\documentclass[runningheads,a4paper]{cmmr2023}
\usepackage{times}
\usepackage{amssymb}
\usepackage{graphicx}
\usepackage{color}
\usepackage{float}
\usepackage{multirow}
\usepackage{hyperref}
\usepackage{subcaption}
\usepackage[table]{xcolor}
\newcommand{\keywords}[1]{\par\addvspace\baselineskip
\noindent\keywordname\enspace\ignorespaces#1}

\begin{document}


\mainmatter  

\title{ShredGP: Guitarist Style-Conditioned Tablature Generation with Transformers}

\titlerunning{ShredGP: Guitarist Style-Conditioned Tablature Generation}

%
%
\author{Pedro Sarmento\inst{1}\thanks{This work is supported by the EPSRC UKRI Centre for Doctoral Training in Artificial Intelligence and Music (Grant no. EP/S022694/1).}\and Adarsh Kumar\inst{2} \and Dekun Xie\inst{1}\and CJ Carr\inst{3}\and and Zack Zukowski\inst{3}\and Mathieu Barthet\inst{1}}
%
\authorrunning{Pedro Sarmento et al.}

\institute{Queen Mary University of London \and Indian Institute of Technology Kharaghpur\ \and Dadabots\\ \email{p.p.sarmento@qmul.ac.uk}}

%

\maketitle


\begin{abstract}

GuitarPro format tablatures are a type of digital music notation that encapsulates information about guitar playing techniques and fingerings. We introduce ShredGP, a GuitarPro tablature generative Transformer-based model conditioned to imitate the style of four distinct iconic electric guitarists. In order to assess the idiosyncrasies of each guitar player, we adopt a computational musicology methodology by analysing features computed from the tokens yielded by the DadaGP encoding scheme. Statistical analyses of the features evidence significant differences between the four guitarists. We trained two variants of the ShredGP model, one using a multi-instrument corpus, the other using solo guitar data. We present a BERT-based model for guitar player classification and use it to evaluate the generated examples. Overall, results from the classifier show that ShredGP is able to generate content congruent with the style of the targeted guitar player. Finally, we reflect on prospective applications for ShredGP for human-AI music interaction.


\end{abstract}

\section{Introduction}

Historically, symbolic music generation research has initially relied on datasets using formats such as MIDI, MusicXML, and ABC \cite{Dong2020}. The publication of the DadaGP dataset \cite{Sarmento2021} has fostered research on \textit{guitar-focused symbolic music generation}, adopting a symbolic format supporting multiple instruments including tablature information for string instruments. The dataset is built-upon the GuitarPro (GP) format, supporting fingering and expressive information specific to fretted string instruments, features not supported by MIDI.  Related works include GTR-CTRL \cite{Sarmento2023}, a Transformer-based generative model for guitar tablatures that can be conditioned on musical genre and instrumentation. In this work, we follow some of the future work suggestions from the GTR-CTRL paper, namely the development of a model conditioned on artist style. As a use case, we selected four guitar players of distinct styles to assess the ability of the model to capture and reproduce their stylistic idiosyncrasies: David Gilmour (DG), Jimi Hendrix (JH), Steve Vai (SV) and Yngwie Malmsteen (YM). Token-based heuristics to analyse the style of the guitarists are presented in Section \ref{cond-sub}, following a computational musicology approach. We present ShredGP, a model that leverages the approach used in GTR-CTRL to condition tablature generation based on guitarist style. The main contributions of this paper are: (1) a method for the generation of multi-instrument guitar tablatures conditioned on artist style; (2) ShredGP, a model for the generation of guitar tablatures in the style of specific guitarists; (3) heuristics to analyze guitar playing styles in the symbolic domain using the token format from DadaGP, that can find applications in computational musicology; (4) SoloGPBERT, a classification model for the task of identifying performances from different guitarists, fine-tuned on the specific four guitar players as a use case.

\section{Background}

\subsection{Deep Learning for Symbolic Music Generation}

The task of music generation with deep learning has been steadily demonstrating promising results and achieving state-of-the-art \cite{Mozer1994}\cite{Eck2002}\cite{Kaliakatsos-Papakostas2020}\cite{sarmentoPerspectivesFutureSonic2021}. Architectural choices for generative music models range from Variational Autoencoders (VAEs) \cite{Roberts2018}\cite{Tan2020}, to Generative Adversarial Networks (GANs) \cite{Dong2018a}\cite{Dong2018}, and natural language processing (NLP) inspired models, such as Recurrent Neural Networks (RNNs) \cite{Meade2019} and, most notably, Transformers \cite{AnnaHuang2019}. This work explores the use of the Transformer, a sequence-to-sequence model capable of learning the dependencies and patterns among elements of a given sequence by incorporating the notion of self-attention, which has achieved state-of-the-art results in many NLP tasks. Huang et al.'s Music Transformer \cite{AnnaHuang2019} was the pioneering work to employ self-attention mechanisms for generating longer sequences of symbolic piano music. Other noteworthy contributions in this area include Musenet \cite{christine_2019}, that used the GPT-2 Transformer model to produce symbolic multi-instrument music across various musical genres; the Pop Music Transformer \cite{Huang2020}, which used the Transformer-XL architecture and demonstrated better rhythmic structure in generating pop piano symbolic music, and the Compound Word Transformer \cite{Hsiao2021}, which explores innovative and more efficient approaches to tokenizing symbolic music during training. 

\subsection{Automatic Guitar Tablature Music Generation}

Despite the widespread availability of guitar tablatures \cite{Macrae2011}\cite{barthet2011}, there has been limited research on generating guitar tablature music prior to the release of the DadaGP dataset \cite{Sarmento2021}. McVicar \cite{McVicar2014} proposed an automatic guitar solo generator in tablature format, which utilized probabilistic models and relied on input chord and key sequences. In terms of guitar tab music generation using deep learning techniques, Chen et al. \cite{Chen2020} developed a fingerstyle guitar generator, trained on a dataset of 333 examples that did not use the GuitarPro format. With the release of the DadaGP dataset \cite{Sarmento2021}, works regarding automatic guitar tablature generation include GTR-CTRL, a Transformer-XL based model that can control instrumentation and musical genre \cite{Sarmento2023}, and LooperGP, a model that can create loops and which was designed having in mind live coding performance applications \cite{Adkins2023}.

\section{Datasets}

\subsection{DadaGP Dataset}

The DadaGP dataset \cite{Sarmento2021} contains a collection of 26,181 songs, available in two different representations: the \textit{token format}, a form of a textual representation of the songs, and the \textit{GuitarPro format}, named after the GuitarPro software used for tablature editing and playback. The conversion between these two file formats is facilitated by a tool that uses PyGuitarPro \cite{PyguitarPro}, a Python library that can parse GuitarPro files. The songs in the DadaGP's \textit{token format} begin with specific tokens such as \verb|artist|, \verb|downtune|, \verb|tempo|, and \verb|start|. Notes played on pitched instruments are represented by tokens in the format \verb|instrument:note:string:fret|. Although this syntax is primarily suitable for string instruments, the combination of string and fret is eventually mapped to a MIDI note, thus supporting other pitched instruments. Percussive instruments, such as the drumkit, are represented using tokens in the form \verb|drums:note:type|. To quantify note durations, the dataset employs the \verb|wait:ticks| token, which uses a resolution of 960 ticks per quarter note. In terms of notating guitar playing techniques, DadaGP uses the note effect (\verb|nfx|) and beat effect (\verb|bfx|) tokens. These include expressive guitar techniques such as \textit{palm mute} (a technique in which the player dampens the strings with their right hand palm), bends and vibratos, tappings,  slides, hammer-ons and pull-offs (both represented under the \verb|nfx:hammer| token).

\subsection{SoloGP Dataset}

In order to create a subset of DadaGP that consisted only of solo guitar parts, we developed a method to extract solo sections from the dataset. By leveraging PyGuitarPro, we searched for \textit{Solo} markers on the files, textual indications of where a guitar solo section is located, then extracted the corresponding guitar part at that section. With this procedure we assembled SoloGP, containing 3,308 guitar solos from more than 1,000 guitarists (12,7\% of the tracks in DadaGP), with a total duration of over than 43h.


\section{Computational Musicology for Guitarist Style Analysis}\label{cond-sub}


In order to experiment with guitarist style-conditioned guitar tablature generation, we gathered a corpus of 50 songs from four distinct iconic electric guitar players: David Gilmour (DG), Jimi Hendrix (JH), Steve Vai (SV) and Yngwie Malmsteen (YM). These guitar players are known to have different styles. To validate this, we use a computational musicology approach \cite{Cook2004} by comparing features computed on a corpus of examples from each guitarist. We use a Type I error $\alpha$ of .05 in statistical analyses. General descriptive statistics about the corpus of each guitar player can be seen in Table \ref{tab:stats}. We can observe, for example, that YM not only plays a higher number of notes than the other three guitar players, but it does it on average at faster tempos. By opposition, DG usually resorts on slower tempos and fewer notes. Additionally, JH and SV seem to make use of specific guitar techniques (e.g. \textit{bends} and \textit{tapping}, respectively) more often, evidenced by a larger number of \verb|nfx| and \verb|bfx| tokens.

\begin{table}[H]
    \centering
    \caption{Overall statistics of the conditioning subset, per guitar player.}
    \vspace{0.2cm}
\setlength{\tabcolsep}{10pt}
\renewcommand{\arraystretch}{1.1}
\scalebox{1}{%
    \begin{tabular}{l|c|c|c}

        \textbf{Guitarist} & \textbf{Avg. Tempo}  & \textbf{Num. Notes} & \textbf{Num. FXs} \\
\hline
\hline
        David Gilmour       & 94 bpm  & 13,534 & 5,921 \\
        Jimi Hendrix        & 111 bpm & 28,843 & 14,625 \\
        Steve Vai           & 123 bpm & 31,715 & 14,457 \\
        Yngwie Malmsteen    & 142 bpm & 33,206 & 7,093 \\
        \hline
        \hline
    \end{tabular}}

    \label{tab:stats}
\end{table}

\begin{figure}[H]
    \centering
    \includegraphics[width=0.9\textwidth]{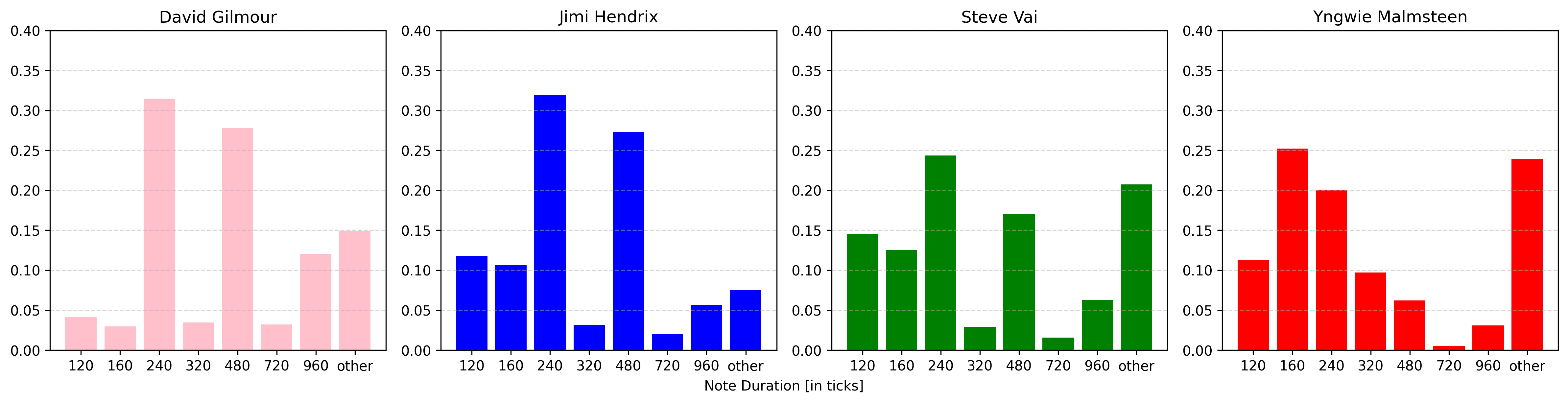}
    \caption{Note duration distribution (in ticks, 960 ticks per quarter note), per guitar player.}
    \label{fig:note-duration-dist}
\end{figure}

Figure \ref{fig:note-duration-dist} presents a distribution of the note durations used by each guitarist in the corpus. The results suggest that, whereas DG and JH seem to predominantly use 16th and 8th notes (240 and 480 ticks, respectively), YM plays 16th note triples more frequently (160 ticks). Furthermore, a Kruskal-Wallis rank sum test yielded significant statistical differences between the four note duration distributions ($H(3)=12.848$, $p<.005$).

\begin{figure}[H]
    \centering
    \includegraphics[width=0.9\textwidth]{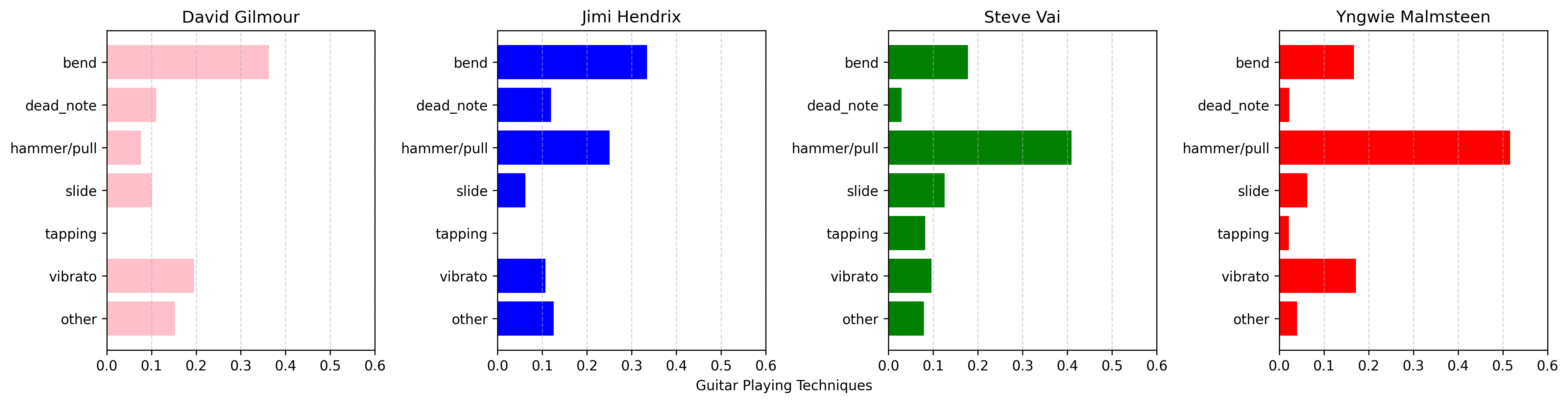}
    \caption{Guitar playing techniques distribution, per guitar player.}
    \label{fig:tech}
\end{figure}

Figure \ref{fig:tech} shows the distributions of guitar playing expressive techniques for each guitar player. An overall analysis indicates that DG relies mostly on \textit{bends} by comparison with the other remaining five techniques. YM seems to prefer \textit{hammer-ons} and \textit{pull-offs} (i.e. left hand \textit{legatos}). It is interesting to note that, for the analyzed corpus, SV is the only guitarist using \textit{tapping} (i.e. a guitar technique in which the player hits a fretted note with a finger from the right hand). A Kruskal-Wallis rank sum test showed significant statistical differences between the four guitar playing techniques' distributions ($H(3)=24.312$, $p<.001$).

In order to have a better melodic/harmonic understanding of each players' performances, we computed the \textbf{pitch class entropy} (PCE) and \textbf{scale consistency} (SC) metrics, as defined by \cite{Dong2020}. Applied to tonal music, PCE can indicate indirectly how tonal a piece is. Applied to a corpus of tonal music, it reflects the consistency in the keys used. The SC is defined as the largest pitch-in-scale rate over all major and minor scales.

\begin{figure}[H]
    \centering
    \includegraphics[width=0.75\textwidth]{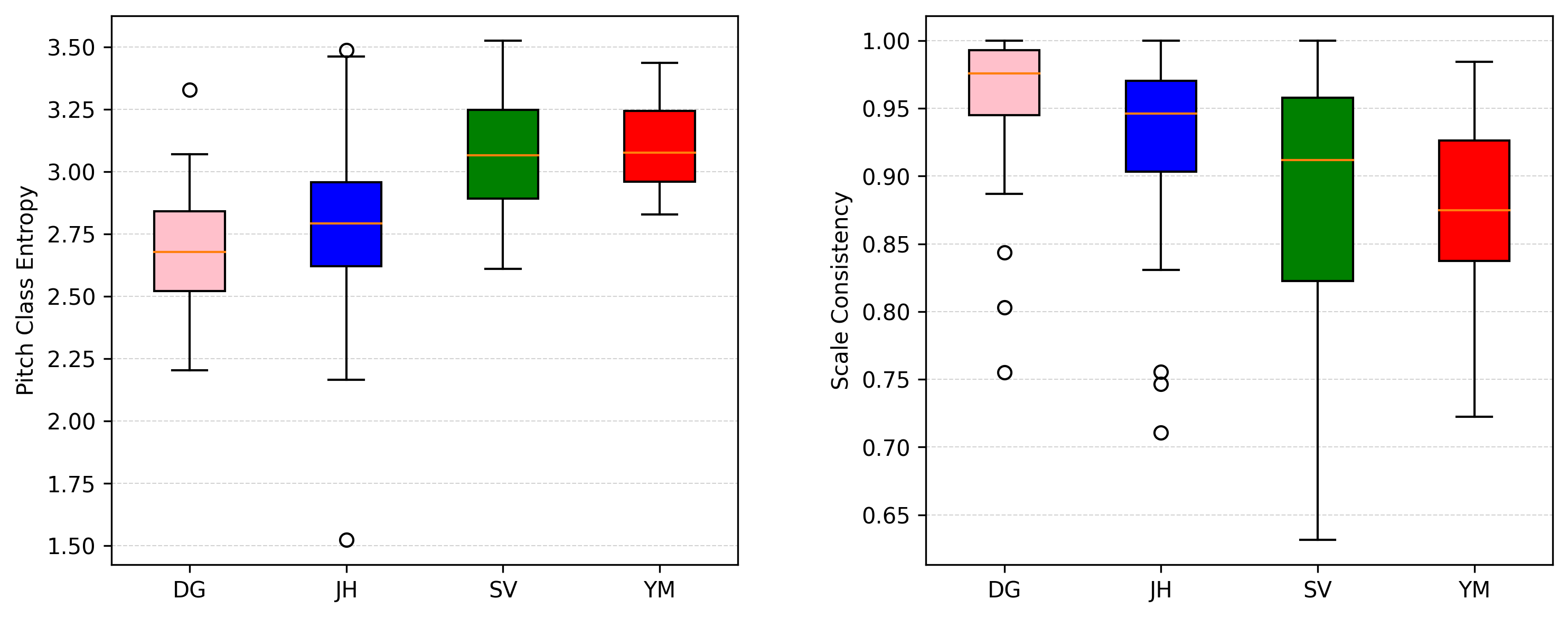}
    \caption{Box plots for pitch class entropy (left) and scale consistency (right) per guitar player.}
    \label{fig:PCE-SC}
\end{figure}

Both plots in Figure \ref{fig:PCE-SC} suggest that SV and YM are more diverse in terms of the keys and pitches used. For PCE, a a Kruskal-Wallis rank sum test showed significant statistical differences between the four guitar players ($H(3)=73.602$, $p<.001$). Likewise, a Kruskal-Wallis rank sum test yielded significant statistical differences between the four guitarists' SC distributions ($H(3)=915.960$, $p<.001$).

 

 

\begin{figure}[H]
    \centering
    \includegraphics[width=0.95\textwidth]{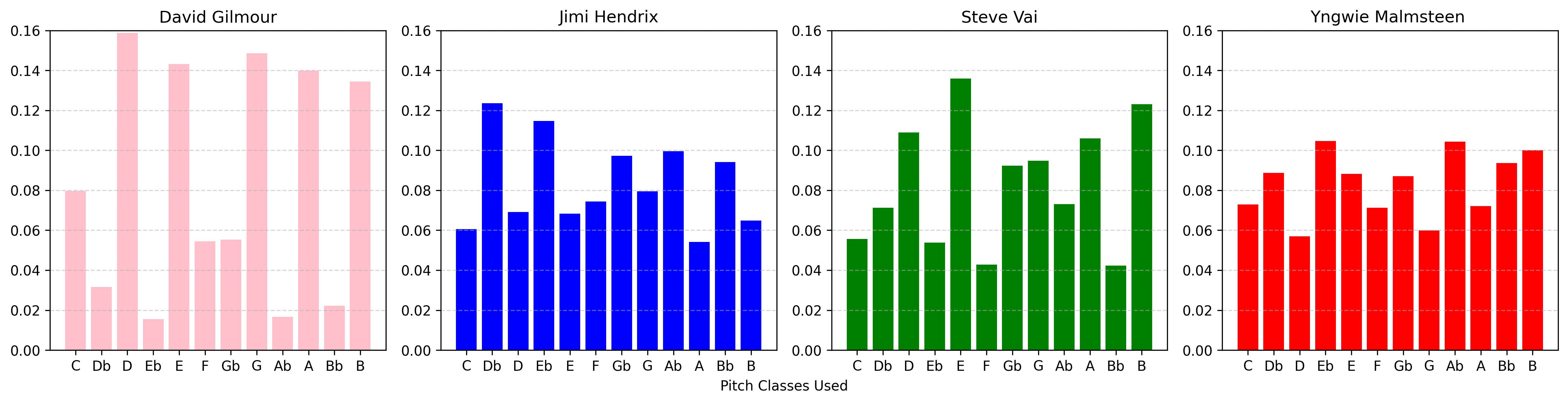}
    \caption{Distributions of pitch classes used, per guitar player.}
    \label{fig:pitches-used}
\end{figure}

Some additional information about the results from PCE and SC can be observed in Figure \ref{fig:pitches-used}. For example, by analyzing the plot for JH, we notice that the five pitch class peaks could correspond to a \textit{Eb} minor pentatonic scale (i.e. \textit{Eb}, \textit{Gb}, \textit{Ab}, \textit{Bb}, \textit{Db}). This is particularly relevant because JH mostly plays in a half-step down guitar tuning (i.e. from the lowest to the highest string: \textit{Eb}-\textit{Ab}-\textit{Db}-\textit{Gb}-\textit{Bb}-\textit{Eb}) and is famous for his use of the minor pentatonic scale. Regarding YM, the other guitarist from the corpus that plays with a half-step down tuning, we can observe that the highest peak also falls on \textit{Eb}. Finally, DG's distribution has visibly lower entropy than the others, in accordance with the plots in Figure \ref{fig:PCE-SC}.

\section{Experiments}


Previous work in \cite{Sarmento2021} demonstrated that the use of \textit{control tokens} succeeded in conditioning a guitar tablature generation model on either instrumentation or musical genre. Following a similar approach, we used the \verb|artist| tokens at the beginning of every song to condition the generation on the style of the four guitarists. Although we investigate here a use case for guitarist style imitation, our approach can be used to condition generation for a whole band (e.g. Pink Floyd). Thus, at the start of every respective song we used a \textit{control token} in the form of \verb|artist:pink_floyd| for DG, a \textit{control token} in the form of \verb|artist:pink_floyd| for DG, a \textit{control token} in the form of  \verb|artist:jimi_hendrix| for JH, a \textit{control token} in the form of  \verb|artist:steve_vai| for SV and  a \verb|artist:yngwie_malmsteen| \textit{control token} for YM.
By training the model with these control tokens, we aim to stir the generation output at the time of inference. In order to obtain varied generated songs, we followed two distinct strategies, one using \textbf{m}ulti-instrument compositions (ShredGP-\textbf{M}), trained on the DadaGP dataset and fine-tuned on the multi-instrument version of the conditioning subset, and another using only \textbf{s}olo-instrument parts (ShredGP-\textbf{S}), trained on the SoloGP dataset and fine-tuned on a solo-instrument version of the conditioning subset (i.e. we manually filtered the guitar parts of every song in this corpus).

\subsection{Model Description}\label{model-desc}

Regarding architectural choices, we followed a similar procedure as in our previous work \cite{Sarmento2021}, namely a Transformer-XL model \cite{Dai} as backbone architecture. Concerning ShredGP-M,  the model's configuration comprised 12 self-attention layers with 8 multi-attention heads, trained for 200 epochs on the whole DadaGP dataset, and fine-tuned for 20 epochs on the multi-instrument conditioning set, with a learning rate of $1e-4$ and a batch size of 8 samples. Regarding the ShredGP-S model, due to the lesser complexity of the task (i.e. generative procedure for a single instrument vs. many instruments), we reduced the models' complexity to  2 self-attention layers and 4 multi-attention heads. ShredGP-S was trained on the SoloGP dataset for 300 epochs, and finally fine-tuned on the conditioning subset for 200 epochs. Both models were training using NVIDIA QUADRO RTX 600 GPUs. 
Model parameters were heuristically tuned based on prior experiments.

\subsection{Inference Procedures}\label{inf-proc}

The results from GTR-CTRL \cite{Sarmento2023} showed a significant effect of the prompting strategy in conditioning guitar tablature generative models on instrumentation and musical genre. As another source of variability for generated outputs, we also use two distinct prompts for both the ShredGP-M and ShredGP-S models: a \textit{full-prompt}, consisting of the first two measures from compositions of the target guitarist, and an \textit{empty-prompt}, comprising only one initial note. In the case of the \textit{full-prompt} for ShredGP (ShredGP-M-FP), we used a multi-instrument version of said two measures, and used a single-instrument version for the ShredGP-S case (ShredGP-S-FP). A similar reasoning was followed for the \textit{empty-prompt} on both ShredGP-M (ShredGP-M-EP) and ShredGP-S (ShredGP-S-EP). We generated 400 examples per model/prompt configuration, comprising a total of 100 songs per guitar player. For ShredGP-S we defined a limit of 256 generated tokens per song, and for ShredGP-M a limit of 2,048 tokens, as ShredGP-M was set to generate multi-instrument compositions, thus needing more tokens to accommodate for that factor. 

\subsection{Listening Examples}
For the experiment settings described in section \ref{inf-proc}, we cherry-picked examples of generated songs for each guitar player. These examples, together with all the generated compositions, without any post-processing, are made available for listening\footnote{Currently available at: \sloppy\url{https://drive.google.com/drive/folders/1vmaKGYFgp-02fGuEvz9BXtWDZuvHk0Hc?usp=share_link}}.

\section{Objective Analysis}\label{obj-anal}


Assessing the quality of generative music models is a difficult task, as it usually involves conducting subjective listening tests that are challenging to design and require significant expertise and resources.
For the particular case of this study, a listening test would need participants that are familiar with the differences in playing style of the four guitarists. Thus, we resorted on an objective computational analysis based on the metrics described in the next subsection. Finally, we compared these results against the ones obtained for the groundtruth data and presented on Section \ref{cond-sub}. 

\subsection{Metrics}


\hspace{\parindent}\textbf{Note Duration Distributions:} we calculated note duration distributions on the generated corpus. We computed the Kullback-Leibler divergence (KLD) between the note duration distributions of the generated examples and of the groundtruth data to assess the similarities between these sets. Here, a smaller value indicates less divergences, hence more similarity. \textbf{Guitar Playing Techniques Distributions:} we computed these distributions for every guitarist/prompt configuration and calculated the KLD between the groundtruth and generated examples. \textbf{SoloGPBERT Classifier:} inspired by the work in GTR-CTRL \cite{Sarmento2023}, we here propose SoloGPBERT, a variant of the model introduced in \cite{Chou2021}, MIDIBERT, as a Bidirectional Encoder Representations from Transformers (BERT)-based masked language able to be configured for downstream classification tasks concerning piano MIDI songs. For SoloGPBERT, we first pre-trained it on the SoloGP dataset for 50 epochs, finally fine-tuning it for two epochs on the conditioning subset for the task of classifying songs of each of the four guitar players, with a split of 55/20/25 between training, validation and test sets. After the fine-tuning, we obtained an accuracy of 89.09\% on the test data, thus deeming this model suitable to distinguish the style of each guitarist with a high confidence.


\subsection{Results}

Results for the Kullback-Leibler divergence (KLD) figures for both the note duration and guitar playing techniques' distributions can be seen in Table \ref{ND-GPT}. 

\begin{figure}[H]
  \centering
  \captionof{table}{KLD scores between the groundtruth distributions for the note duration (left) and guitar playing techniques (right) and the distributions for the generations for each of the four guitar players. Best results in \textbf{bold}.} 
  \label{ND-GPT}

  \begin{subfigure}[b]{0.45\textwidth}
    \centering
    \scalebox{0.80}{%
      \begin{tabular}{l|c|c|c|c|c}
\multicolumn{1}{l}{}       &     & \multicolumn{4}{c}{\textbf{Note Durations}}                                        \\ 
\cline{3-6}
\multicolumn{1}{l}{}       &     & \textbf{DG}           & \textbf{JH}            & \textbf{SV}            & \textbf{YM}        \\ 
\hline
\multirow{4}{*}{\rotatebox[origin=c]{90}{\textbf{DG}}}
                           & M-FP & \cellcolor{pink!25}  0.2808 &  \textbf{0.2051}          &  0.0797          &  0.3045     \\
                           & M-EP & \cellcolor{pink!25}  \textbf{0.0975} &  0.2610          &  0.1435          &  0.3963     \\
                           & S-FP & \cellcolor{pink!25}  \textbf{0.0497} &  0.2575          &  0.1503          &  0.4455     \\
                           & S-EP & \cellcolor{pink!25}  \textbf{0.0546} &  0.0990          &  0.2775          &  0.7579     \\ 
\hline
\multirow{4}{*}{\rotatebox[origin=c]{90}{\textbf{JH}}}    
                           & M-FP &  \textbf{0.0877} &   \cellcolor{blue!25} 0.2160         &  0.3442          &  0.9227     \\
                           & M-EP &  0.2721 &   \cellcolor{blue!25} 0.5388         &  \textbf{0.2050}          &  0.3859     \\
                           & S-FP &  \textbf{0.2133} &   \cellcolor{blue!25} 0.2805         &  0.4967          &  1.1245     \\
                           & S-EP &  0.2300 &   \cellcolor{blue!25} \textbf{0.2120}         &  0.4862          &  1.094     \\ 
\hline
\multirow{4}{*}{\rotatebox[origin=c]{90}{\textbf{SV}}}    
                           & M-FP &  0.1542 &   0.1998      &  \cellcolor{green!25} \textbf{0.1114}       &  0.3814          \\
                           & M-EP &  0.0920 &   0.1465      &  \cellcolor{green!25} \textbf{0.0884}       &  0.3844          \\
                           & S-FP &  0.2263 &   \textbf{0.1790}      &  \cellcolor{green!25} 0.3727       &  0.6814          \\
                           & S-EP &  0.2343 &   \textbf{0.0582}      &  \cellcolor{green!25} 0.1428       &  0.5225          \\ 
\hline
\multirow{4}{*}{\rotatebox[origin=c]{90}{\textbf{YM}}} 
                           & M-FP &  1.4008 &   1.3423      &  \textbf{0.7131}       &  \cellcolor{red!25} 0.7169          \\
                           & M-EP &  0.9753 &   0.7833      &  0.3224       &  \cellcolor{red!25} \textbf{0.2919}          \\
                           & S-FP &  1.3478 &   0.6474      &  0.4610       &  \cellcolor{red!25} \textbf{0.4452}          \\
                           & S-EP &  1.6598 &   0.8734      &  0.7884       &  \cellcolor{red!25} \textbf{0.4136}          \\ 
\hline

\end{tabular}%
}
  \end{subfigure}
  \hfill
  \begin{subfigure}[b]{0.45\textwidth}
    \centering
    \scalebox{0.80}{%
      \begin{tabular}{l|c|c|c|c|c}
\multicolumn{1}{l}{}       &     & \multicolumn{4}{c}{\textbf{Guitar Playing Techniques}}                                        \\ 
\cline{3-6}
\multicolumn{1}{l}{}       &     & \textbf{DG}           & \textbf{JH}            & \textbf{SV}            & \textbf{YM}        \\ 
\hline
\multirow{4}{*}{\rotatebox[origin=c]{90}{\textbf{DG}}}
                           & M-FP & \cellcolor{pink!25}  0.3077 &  \textbf{0.0552}          &  0.2305          &  0.2076     \\
                           & M-EP & \cellcolor{pink!25}  0.2108 &  \textbf{0.0875}          &  0.2511          &  0.2008     \\
                           & S-FP & \cellcolor{pink!25}  \textbf{0.1497} &  0.2952          &  0.6834          &  0.6269     \\
                           & S-EP & \cellcolor{pink!25}  \textbf{0.1239} &  0.2832          &  0.5308          &  0.4330     \\ 
\hline
\multirow{4}{*}{\rotatebox[origin=c]{90}{\textbf{JH}}}    
                           & M-FP &  \textbf{0.3052} &   \cellcolor{blue!25} 0.3159        &  0.7550          &  0.8674    \\
                           & M-EP &  0.05351 &   \cellcolor{blue!25} \textbf{0.2849}        &  0.6000          &  0.6267     \\
                           & S-FP &  0.1683 &   \cellcolor{blue!25} \textbf{0.0990}         &  0.3063          &  0.3687     \\
                           & S-EP &  0.1618    &   \cellcolor{blue!25} \textbf{0.0933}         &  0.3707          &  0.3268     \\ 
\hline
\multirow{4}{*}{\rotatebox[origin=c]{90}{\textbf{SV}}}    
                           & M-FP &  0.3053    &   \textbf{0.1121}      &  \cellcolor{green!25} 0.2891       &  0.4883          \\
                           & M-EP &  0.7508 &   \textbf{0.2293}      &  \cellcolor{green!25} 0.2924       &  0.3234          \\
                           & S-FP &  \textbf{0.2601}    &   0.3232      &  \cellcolor{green!25} 0.4413       &  0.7312          \\
                           & S-EP &  \textbf{0.2415} &   0.2656      &  \cellcolor{green!25} 0.3217       &  0.2628          \\ 
\hline
\multirow{4}{*}{\rotatebox[origin=c]{90}{\textbf{YM}}} 
                           & M-FP &  0.9105    &   0.3299      &  0.1876       &  \cellcolor{red!25} \textbf{0.0449}          \\
                           & M-EP &  1.3512    &   0.5029      &  0.2511       &  \cellcolor{red!25} \textbf{0.1439}          \\
                           & S-FP &  0.3894    &   0.3820      &  0.5410       &  \cellcolor{red!25} \textbf{0.3192}          \\
                           & S-EP &  1.9688   &   0.8898      &  0.4014       &  \cellcolor{red!25} \textbf{0.2360}          \\ 
\hline

\end{tabular}%
}

  \end{subfigure}
  \end{figure}
  

Concerning \textbf{note duration}'s distributions (left side table), the generative outputs conditioned on DG and YM seem to obtained the best classifications (i.e. 3 best classifications out of 4 possible model/prompt configurations). Generating compositions with a note duration distribution similar to the groundtruth from JH obtains the worst scores (i.e. only 1 best classifications out of 4 possible). Regarding the \textbf{guitar playing techniques}' distributions (right side table), YM obtained the best results, while SV-conditioned generations failed to match the groundtruth distribution. Considering the figures of both tables together, an overall analysis suggests that the style from SV is the hardest to model (2/8 best results), whilst YM obtains the highest number of best scores (7/8 best results), followed by DG (5/8) and JH (4/8).

\begin{table}[H]
 \caption{Guitar player classification softmax scores from SoloGPBERT, for the generations from every guitarist/prompt configuration. Best results in \textbf{bold}.}
    \vspace{0.1cm}
\centering
\setlength{\arrayrulewidth}{1pt}
\arrayrulecolor{black}
\setlength{\tabcolsep}{6pt}
\renewcommand{\arraystretch}{1.1}
\scalebox{0.80}{%
\begin{tabular}{l|c|c|c|c|c}
\multicolumn{1}{l}{}       &     & \multicolumn{4}{c}{\textbf{Guitar Player Classification Score}}                                        \\ 
\cline{3-6}
\multicolumn{1}{l}{}       &     & \textbf{DG}           & \textbf{JH}            & \textbf{SV}            & \textbf{YM}        \\ 
\hline
\multirow{4}{*}{\rotatebox[origin=c]{90}{\textbf{DG}}}
                           & M-FP & \cellcolor{pink!25}  \textbf{0.5691} &  0.2103          &  0.1175          &  0.1031     \\
                           & M-EP & \cellcolor{pink!25}  \textbf{0.5086} &  0.2033          &  0.1474          &  0.1407     \\
                           & S-FP & \cellcolor{pink!25}  \textbf{0.6037} &  0.2238          &  0.0945          &  0.0780     \\
                           & S-EP & \cellcolor{pink!25}  \textbf{0.5952} &  0.2034          &  0.1080          &  0.1006     \\ 
\hline
\multirow{4}{*}{\rotatebox[origin=c]{90}{\textbf{JH}}}    
                           & M-FP &  0.1577 &   \cellcolor{blue!25} \textbf{0.5785}         &  0.1358          &  0.1280     \\
                           & M-EP &  0.2850 &   \cellcolor{blue!25} \textbf{0.4054}         &  0.1338          &  0.1757     \\
                           & S-FP &  0.1229 &   \cellcolor{blue!25} \textbf{0.6285}         &  0.1105          &  0.1380     \\
                           & S-EP &  0.1090 &   \cellcolor{blue!25} \textbf{0.6207}         &  0.0835          &  0.1868     \\ 
\hline
\multirow{4}{*}{\rotatebox[origin=c]{90}{\textbf{SV}}}    
                           & M-FP &  0.1839 &   0.3146      &  \cellcolor{green!25} \textbf{0.3318}       &  0.1697          \\
                           & M-EP &  0.1859 &   0.2146      &  \cellcolor{green!25} \textbf{0.3498}       &  0.2497          \\
                           & S-FP &  0.1692 &   \textbf{0.3499}      &  \cellcolor{green!25} 0.3042       &  0.1768          \\
                           & S-EP &  0.0619 &   0.2831      &  \cellcolor{green!25} \textbf{0.3273}       &  0.2827          \\ 
\hline
\multirow{4}{*}{\rotatebox[origin=c]{90}{\textbf{YM}}} 
                           & M-FP &  0.0848 &   0.2364      &  0.0979       &  \cellcolor{red!25} \textbf{0.5810}          \\
                           & M-EP &  0.1085 &   0.2512      &  0.1161       &  \cellcolor{red!25} \textbf{0.5242}          \\
                           & S-FP &  0.0619 &   0.2156      &  0.0755       &  \cellcolor{red!25} \textbf{0.6470}          \\
                           & S-EP &  0.0461 &   0.1501      &  0.0557       &  \cellcolor{red!25} \textbf{0.7480}          \\ 
\hline

\end{tabular}%
}
   \label{tab-sologpbert}
\end{table}

The results obtained from the SoloGPBERT classifier can be observed in Table \ref{tab-sologpbert}. Overall, the generations from all guitarist/prompt configurations were classified correctly, with the exception of ShredGP-S-FP when conditioned on SV, thus showcasing ShredGP's ability to recreate compositions on the style of each guitarist. It is interesting to note that this matches the conclusions from the results in the previous metrics, where SV also proved to be harder to model. Similarly, the results for YM obtain the best classifications on all the prompt configurations.

\section{Subjective Analysis}

\begin{figure}[H]
    \centering
    \includegraphics[width=0.77\textwidth]{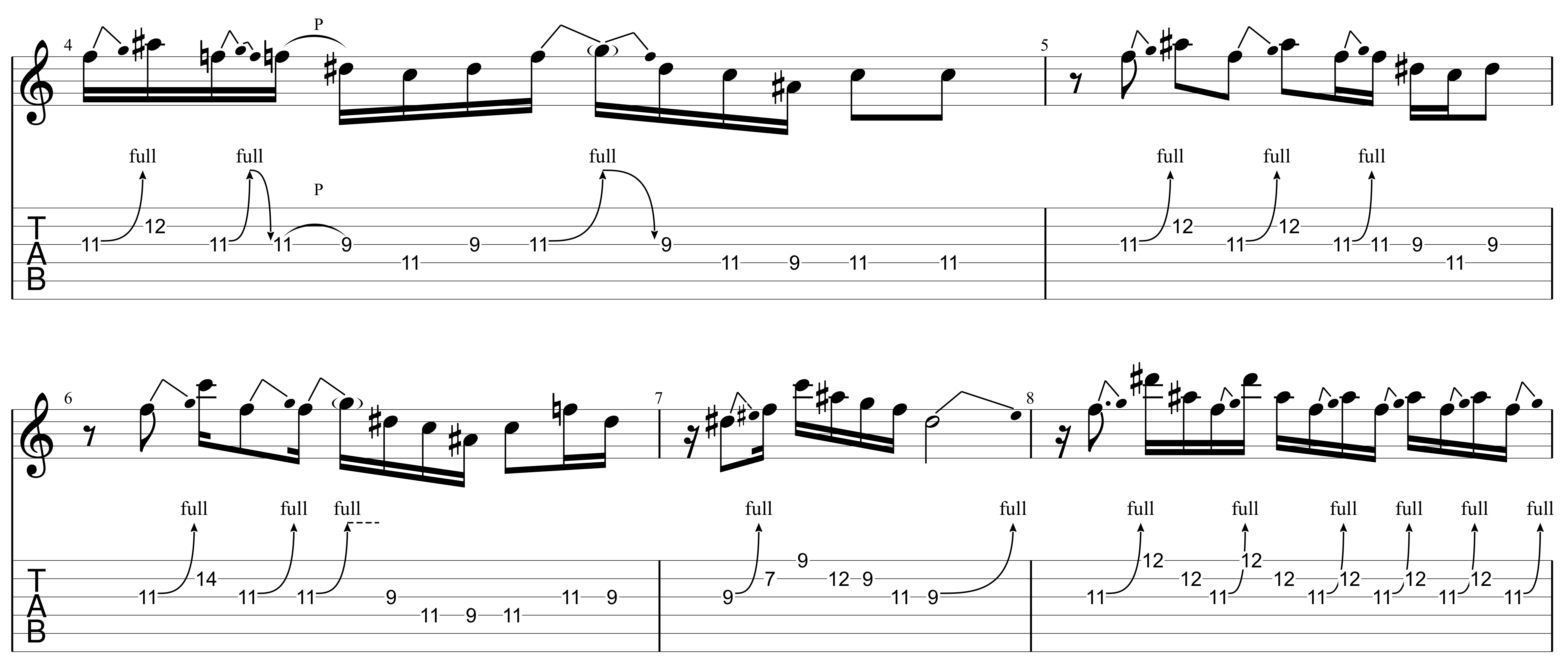}
    \caption{Measures 4 to 8 in \href{https://drive.google.com/file/d/1FLBNzTG6Z0XU87s6n4Kz0D6EyP0xtfGY/view?usp=share_link}{\underline{ShredGP-M-FP id-003 from JH}}, in 4/4.}
    \label{fig:jh}
\end{figure}

In order to complement the quantitative evaluation, we conducted a subjective analysis of some of the cherry-picked examples. In this section, underlinked song ids in figures' captions are hyperlinked to facilitate listening. We would like to highlight that what we define as \textit{style} in this paper is viewed from the perspective of a symbolic representation of these guitarists' playing techniques, thus not taking into account timbral features that express identifiable, unique characteristics of each guitar player. In Figure \ref{fig:jh} we display a few measures from a song from ShredGP-M-FP conditioned on JH. We can observe a stylistic phrasing that emphasizes the minor pentatonic scale, composed of patterns with bends that are characteristic of JH.

\begin{figure}[H]
    \centering
    \includegraphics[width=0.77\textwidth]{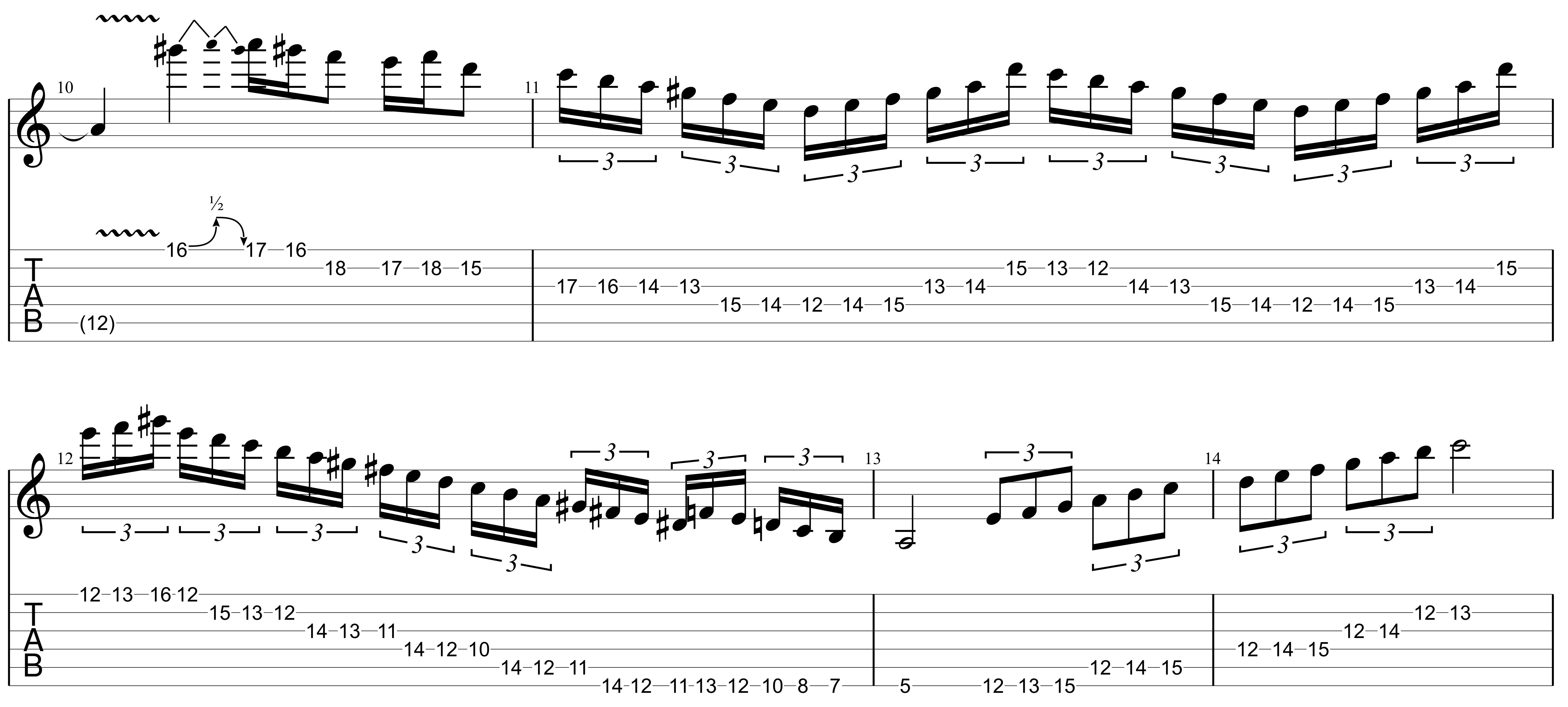}
    \caption{Measures 10 to 14 in \href{https://drive.google.com/file/d/1c9PdUOCwfzBFanVK1mEy_BsS3iQfUZwS/view?usp=share_link}{\underline{ShredGP-M-EP id-002 from YM}}, in 4/4.}
    \label{fig:ym}
\end{figure}

Similarly, the style of YM is also visible in Figure \ref{fig:ym}, an example from ShredGP-M-EP. The fast 16th note triplet patterns can be observed in measures 11 and 12, supporting the findings on Figure \ref{fig:note-duration-dist} that show that YM mostly resorts on this type of note durations.

\section{Discussion}

Overall, ShredGP succeeds on imitating the style of the four guitarists, its results supported by both an objective and subjective analysis. Regarding the evaluation procedures presented in Section \ref{obj-anal}, we would like to clarify that the development of both ShredGP-M and ShredGP-S in parallel was not carried out with the intent of comparing the two approaches, but only to provide some variety in terms of the procedures. Ultimately, we intended to investigate if we could create a model conditioned on guitarist playing styles. Furthermore, using an AI model (SoloGPBERT) to evaluate the results of another AI model (ShredGP) presents limitations, but due to difficulties in recruiting expert participants, we deem that it can provide a preliminary assessment, as it showed promising results in previous work \cite{Sarmento2023}. Moreover, reflecting on some of the findings from the computational musicology analysis outlined in Section \ref{cond-sub}, we believe that the token format in DadaGP opens up new possibilities for the assessment of guitarist playing styles. We anticipate that applying these heuristics to a wider corpus of guitar players could potentially lead to the creation of {continuous space of guitar playing style}, classifying different guitarists and positioning them in said space accordingly. However, it's worth noticing that these methods do not account for a disentanglement of the guitarist style from the style of the group/band they are playing in, as many times the composition will put creative constraints on the guitar players' part. For our particular case, while JH, SV and YM are theoretically the lead composers in their own groups, the same cannot be said about DG and Pink Floyd. 




\section{Conclusion and Future Work}

In this paper we presented ShredGP, a Transformer-based model for guitar tablature generation, conditioned on the style of four distinct iconic electric guitarists. Furthermore, in order to justify the choice of these guitar players as a conditioning subset, we proposed and implemented a computational musicology-driven approach that leverages DadaGP's token format to analyze guitar players' style on different aspects. Generative outcomes from ShredGP were overall able to match the style of each guitarist, obtaining better results when modelling the guitar playing of YM and worst results for SV. These conclusions are supported by both the SoloGPBERT classifier analysis and the comparison of note duration and guitar playing techniques distributions agaisnt the groundtruth data. In future work we plan to expand the musicological analysis on a wider selection of artists. Regarding the evaluation of our generative results, we expect to better support these findings with listening tests targetting expert guitar players. Finally, we aim to use the methods in this paper for human-AI co-creative collaborations with guitar players.

\bibliographystyle{splncs04}
\bibliography{mybibliography}

\end{document}